\documentclass[11pt]{amsart}

\usepackage{placeins}
\usepackage{amsthm,amsmath,amssymb,url}
\usepackage[top=35mm, bottom=30mm, left=20mm, right=20mm]{geometry}
\usepackage{hyperref}
\usepackage{graphicx}
\usepackage{float}
\usepackage{booktabs}
\usepackage{tikz}
\usepackage{calc}
\usepackage{changepage}
\usepackage{lipsum}
\usepackage{svg}
\usepackage{pifont}
\newcommand{\cmark}{\ding{51}}%
\newcommand{\xmark}{\ding{55}}%

\begin{document} 
\title{Non-cumulative  measures of researcher citation impact} 
\author{Mark C. Wilson}
\address{Department of Mathematics and Statistics, University of Massachusetts Amherst}
\author{Zhou Tang}
\address{Department of Mathematics and Statistics, University of Massachusetts Amherst}
\date{\today}
\begin{abstract}
The most commonly used publication metrics for individual researchers are the the total number of publications, the total number of citations, and Hirsch's $h$-index. Each of these is cumulative, and hence increases throughout a researcher's career, making it less suitable for evaluation of junior researchers or assessing recent impact. Most other author-level measures in the literature share this cumulative property. By contrast, we aim to study non-cumulative measures that answer the question ``in terms of citation impact, what have you done lately?"

We single out six measures from the rather sparse literature, including Hirsch's $m$-index, a time-scaled version of the $h$-index. We introduce new measures based on the idea of ``citation acceleration". After presenting several axioms for non-cumulative measures, we conclude that one of our new measures has much better theoretical justification. We present a small-scale study of its performance on real data and conclude that it shows substantial promise for future use.
\end{abstract}

\date{\today}
\maketitle
\section{Introduction} \label{s:intro}
Despite strong opinion to the contrary among researchers, it is deemed necessary by bureaucrats worldwide to use simple measures of researcher impact. Measures based on research publications (mostly research monographs and peer reviewed articles) are heavily used, the most common being the cumulative number of citations $N(t)$, cumulative number of papers $P(t)$, and the $h$-index $h(t)$ \cite{hirsch2005index} (defined as the greatest integer $h$ such that the author has at least $h$ papers each of which has at least $h$ citations). All three quantities above are biased toward senior scholars, being cumulative and therefore automatically increasing over time, even after the end of the researcher's career. Also, they provide information on overall career citation impact, but no answer to ``what have you done  lately?". For many purposes it is not particularly useful to know the $h$-index of Isaac Newton or total number of citations of Albert Einstein. Comparing researchers near the start of their careers, comparing them with more senior researchers, or trying to predict future productivity and impact of a researcher, clearly require different metrics.

Citation metrics that are not automatically increasing have received much less discussion in the literature. For example, the 
survey by Wildgaard, Schneider and Larsen \cite{wildgaard2014review} of 108 author-level metrics contains at most 15 that are not automatically increasing and which attempt to measure time-varying performance. The earlier survey by Bornmann, Mutz, Hug and Daniel \cite{bornmann2011multilevel} of variants of the $h$-index included only 6 variants that attempted to adjust for career age, out of 37 indicators. Of course, non-increasing measures intended to account for career age have been covered by some researchers. Indeed, Hirsch in his original paper \cite{hirsch2005index} devoted substantial analysis to the rate of growth of $h$ with the number of years $t$ since the author's first publication, and defined the $m$-index by $m(t) = h(t)/t$. He calculated $m$ for a selection of physicists (using a single fixed year, presumably 2005) and concluded that a value of $m$ around $1, 2, 3$ correlated with his judgment of distinguishing between ``successful scientist", ``outstanding scientist" and ``truly unique" individual respectively.  The $m$-index was immediately studied by others \cite{liang2006h, burrell2007hirsch} but has been relatively little explored since. Another measure based on the $h$-index and attempting to measure recent performance is the \emph{contemporary} $h$-index of Sidiropoulos, Katsaros and Manolopoulos \cite{sidiropoulos2007generalized}. Other measures based on the $h$-index and attempting to adjust for career age include the \emph{AR-index} \cite{jin2007r}, which measures the average number of citations per publication, but restricted to publications in the \emph{$h$-core} (the minimal set needed for computation of the $h$-index). The literature contains fewer nonincreasing measures not involving the $h$-index. One could of course also look at the analogue of the AR-index where all publications are considered: the average number of citations per paper. We stop here, conscious that we must draw a line somewhere --- given the axiomatic approach to be taken below, it already seems clear that most of the above measures will fail many of the axioms.

\subsection{Our contribution}
\label{ss:our_contrib}
We argue that the ``instantaneous rate of accumulation of citations owing to recent work" is the relevant measure of recent citation productivity. We claim that this is precisely the ``citation acceleration", the second time-derivative of the number of citations accumulated by an author. Since this quantity is not directly observable, we explore measures aimed at approximating it.

These measures are:
\begin{itemize}
    \item $$w(t):=\frac{2N(t)}{t^2}$$
    \item $$W(t):= N(t)-2N(t-1)+N(t-2)$$
    \item $$W_5(t):= \frac{1}{7} \left(2N(t) - N(t-1) - 2N(t-2) - N(t-3) + 2N(t-4) \right).$$
\end{itemize}

We single out 6 existing measures from the literature that are non-cumulative and intended to adjust for career age.  These are   
\begin{itemize}
    \item Hirsch's \cite{hirsch2005index} $m$-index defined by $$m(t):= h(t)/t$$
    \item Mannella and Rossi's \cite{mannella2013time} measure $$\alpha_1(t):= \frac{h(t)}{\sqrt{N(t)}}$$
    \item The contemporary $h$-index $h^c(t)$ \cite{sidiropoulos2007generalized}, defined below.
    \item The trend $h$-index $h^t(t)$ \cite{sidiropoulos2007generalized}, defined below.
    \item The age-weighted citation rate $A(t)$, defined below.
    \item The average number of citations per year $\mu(t)$, defined below.
\end{itemize}

In Section~\ref{s:axiom} we evaluate all 9 citation measures against axiomatic criteria. The difference between ``theoretical" and ``empirical" work in bibliometrics has been well described by Waltmann and Van Eck \cite{waltman2012inconsistency}. Our approach here is grounded in a theoretical analysis. The axiomatic approach to bibliometric indicators has been applied to the $h$-index by several authors  following the initial paper of Woeginger \cite{woeginger2008axiomatic} --- we single out work by Quesada \cite{quesada2011further} and Bouyssou and Marchant \cite{bouyssou2013interpretable}. To our knowledge, axiomatics have been applied to  few other indicators. We single out work by Waltman and van Eck \cite{waltman2009taxonomy} and by Bouyssou and Marchant  \cite{bouyssou2014axiomatic}, which gives axiomatic characterizations of several well-known indicators such as $N$ and $P$. 

To check that we have not strayed too far from reality, we evaluate the performance of the theoretically best measures $W$ and $W_5$ on several small datasets of mathematical researchers. Results as shown in Section~\ref{ss:results} are promising with respect to  predictive value and  correlation with expert judgment.

\subsubsection*{Definition of remaining measures}

The \emph{age-weighted citation rate} is obtained by summing over all publications the average number of citations per year. The measure $\mu(t)$ is simply the average number of citations per year, evaluated at time $t$.

The contemporary $h$-index is defined analogously to the $h$-index, where citations  gradually lose value.  The contemporary $h$-index uses an adjusted weight of citation, and requires the researcher to have published at least $h$ papers, each with adjusted citation weight to be at least $h$. The adjusted citation weight of a paper published in year $t_0$ and having $C$ citations up to year $t$ is given by 
$$
\gamma \left( t - t_0 + 1\right)^{-\delta} C
$$
where $\delta$ and $\gamma$ are fixed positive constants. For example if $\gamma = \delta = 1$, the adjusted citation weight of a paper is the number of citations per year since the year before its publication.

The trend $h$-index is defined using similar reasoning to the contemporary $h$-index, except that the weight depends on the year of citation, not of publication. Each citing paper in year $t$ contributes a weight of 
$$
\gamma \left( t - t_0 + 1\right)^{-\delta}.
$$

\section{Background and definitions} \label{s:def}
Consider a researcher emitting research publications starting from time $t = 0$. These publications accumulate citations at a certain rate, dependent both on such factors as the number of publications, the size of the research field, the citation practices of the field, the attractiveness of the papers to other researchers. As mentioned above, the most commonly used metrics are: 
\begin{itemize}
\item $P(t)$, the number of publications up to time $t$; 
\item $N(t)$, the number of citations up to time $t$; 
\item $h(t)$, the (Hirsch) $h$-index at time $t$.
\end{itemize}

Note that these each increase in $t$, even after the end of the researcher's career.

\subsection{The simplest citation model} 
\label{ss:simple}
For definiteness we measure time in years and $t = 0$ corresponds to the date of the first publication. We first consider what we call the ``simple model" in continuous time. In this case, a researcher publishes at a constant rate of $p$ papers per unit time and each paper attracts citations at a rate of $c$ per paper per unit time, forever. A discrete version of this model was used by Hirsch \cite{hirsch2005index}. Then the number of publications by time $t$ is $P(t) = pt$, and the total number of citations is
\begin{equation}\label{eq:6}
N(t) = \int_{0}^t pcs \, ds = pct^2/2.
\end{equation}
The ``acceleration" in citation accumulation is therefore
\begin{equation}\label{eq:7}
N''(t) = pc.
\end{equation}

Our main idea is that the quantity $pc$ measures the instantaneous accumulation of citations from new work. This ``acceleration" is a key measure of recent productivity and citation impact. For a given research field, larger values indicate researchers with greater recent impact. 

The instantaneous citation acceleration cannot be measured directly owing to the discreteness of available publication data, so we introduce the quantity
$$
W^{\delta}(t) = \frac{N(t) - 2 N(t - \delta) + N(t-2\delta)}{\delta^2}
$$
which is the usual backward difference approximation to the second derivative. When $\delta = 1$ year, this reduces to the measure $W$ defined above.

We also consider the measure $W_5$, obtained by fitting quadratics to $N$ through successive windows of 5 data points (separated by 1 year intervals) and approximating the second derivative on each. This is an example of a \emph{Savitzky-Golay filter}, widely used to smooth discrete data of this type. In fact $W_5$ is the simplest such filter, and we could define $W_{2k+1}$ analogously for $k>2$ by using larger numbers of points.

The measure $w$ is clearly constant in the simple model, with value $pc$. So is $W^{\delta}$, as shown by
$$
W^{\delta}(t) = \frac{pc}{2} \left(t^2 - 2 (t-\delta)^2 + (t-2\delta)^2 \right) = pc.
$$

So is $W_5$, because the filter used is exact for quadratics and hence reproduces the (constant) second derivative.

As explained by Hirsch \cite{hirsch2005index}, in this model we may assume that the papers in the $h$-core at time $t$ have been published up to time $s\leq t$, so that $ps = h(t)$, while the number of citations of the least cited element of the $h$-core is $c(t-s) = h(t)$. This leads immediately to
$$ h(t)= \frac{pc}{p+c}t  
$$
so that
$$
m(t) = \frac{pc}{p+c}.
$$
As an aside, note that this equals the harmonic mean of $p$ and $c$.

From above it follows that in the simple model
\begin{equation}\label{eq:2}
N(t) = \frac{(p+c)^2}{2pc} h(t)^2
\end{equation}
so that
\begin{equation}\label{eq:3}
h(t)=\beta\sqrt{N(t)} 
\end{equation}
where $\beta = \frac{\sqrt{2pc}}{p+c}$.
This is the reasoning behind the definition of $\alpha_1$, and clearly in the simple model we have
$$
\alpha_1(t) = \frac{\sqrt{2pc}}{p+c}.
$$

Thus in the simple model several measures are constant. However, not all constants are equally valid. The units of acceleration should be $\text{citation}/(\text{year})^2$, and indeed $w, W, W_5$ have these units. However the units of $m$ and $\alpha_1$ are inconsistent with this requirement.

The value of $\mu(t)$ is clearly $pct/2$. The age-weighted citation measure $A(t)$ equals $pct$ in this model, since in \eqref{eq:6} we divide the integrand by $s$. The contemporary $h$-index and trend $h$-index are harder to compute, but are also nonconstant unless $\delta = 1$ (we use the method shown above for the $h$-index, and omit the details here).

\if01 - for another paper
\subsubsection{A more complicated model}
Suppose that each publication by a given researcher attracts the same number of citations, $C(t)$, up to time $t$ after publication. If $d\mu(t)$ is the measure describing the publication rate, then the total number of citations by time $t$ is the convolution
$$
N(t) = \int_{0}^t C(t-s) d\mu(s) = \int_{0}^t C(s) d\mu(t-s) = 
\int_{0}^\infty C(t-s) d\mu(s).
$$
If $d\mu(s) = p ds$ then $N''(t) = C'(t)$. So under a constant production model where the impact of publications varies, we can still 
measure the citation acceleration. In general $N'(t) = C(t) \mu'(t)$. *** check derivative of convolution ***

\fi
\subsection{A model incorporating retirement} \label{ss:second}
We introduce another simple model in order to analyze the prediction value and behavior of measures when a researcher stops publishing papers due to any reason. In the second model we suppose that publications in the simple model stop after time $T$. In  this case, the number of citations is $N(T) =\frac{pcT^2}{2}$, while for $t>T$ we have an additional  $pcT(t-T)$ citations. Thus if $t\geq T$,
\begin{equation}\label{eq:16}
N(t) = pcT\left(t-T/2\right).
\end{equation}

Direct computation shows that the measure $W$ takes the value zero provided $T\geq t+2$, so that sufficient time has elapsed for measurements to be taken. However, the other measures above take on nonzero values when $t>T$. We see that $w$ has the value $pcT(2t-T)/t^2$. For example if $t = 2T$, so the total time elapsed after retirement is as long as the researcher's entire career, $w$ has reduced by only $25\%$, to $3pc/4$, from its previous constant value $pc$ during the career.

Hirsch's original argument shows that the $h$-index has the same value $pct/(p+c)$ in the second model until time $t = T(1+p/c)$ as it had in the simple model, but that after this time it has value $pT/t$. Interestingly, this is independent of $c$. The measure $\alpha_1$ takes the value $\sqrt{\frac{p}{3cT}}$ when $t=2T$. We do not compute the details of the contemporary and trend $h$-index here, since we can see enough to rule them out axiomatically below.

It is of course possible to explore more complicated models, but the focus of 
this article is the introduction of a new measure with axiomatic justification, so we now proceed to that.

\section{Axioms} \label{s:axiom}
As mentioned above, the number of possible measures is enormous, but without axiomatic foundations, it seems pointless to study them in detail. We now present several axioms for measures intended to describe non-cumulative citation impact. All except the last seem to us to be uncontroversial.

\begin{description}
\item[Computability] The measure should be easily computable from citation counts, paper counts, and academic age.
\item[Units] The units of the measure should be $(\text{citation)}/(\text{time})^2$.
\item[Locality] If no citations are gained during a time interval, the measure is zero during that interval.
\item[Constancy] In the first model, the measure is constant.
\item[End of career] In the second model, the measure is zero for $t>T$.
\item[Packaging-independence] The measure should not depend on $P$: it is computable only from citation counts and academic age.

\end{description}

The Packaging-independence axiom requires more explanation. We argue that the impact of a researcher with 10 papers each attracting 100 citations is the same as if all 10 papers had been combined into a book that receives 1000 citations. The packaging into publications may in practice affect the number of citations in a more complicated way, but if publications have no overlap and citations reflect intellectual influence only, this should not occur. Of course, this is also a strong argument against using the $h$-index.

The Locality axiom may need to be interpreted slightly differently when time is discrete. For example, the measure $W$ satisfies this axiom provided the point chosen is 2 years past the left endpoint of the interval in question.

Table~\ref{t:axioms} shows the performance of the abovementioned measures against these axiomatic criteria. Clearly, $W$ and $W_5$ perform much better than the others, and we consider only these measures in the next section. Note that $h^c$ requires specification of two free parameters before we can even compute it, and we know of no principled way to do that, hence the failure of the Computability axiom. The other axioms were evaluated for arbitrary $\gamma$ and $\delta$, and give the same answer for  all choices of these constants (with an exception for the constancy axiom as noted above).

\begin{table}[]
\centering
\caption{Performance of measures with respect to axioms}
\label{t:axioms}
\begin{tabular}{llllllllll}
\hline
\textbf{Axiom$\backslash$ Measure}  & $w$ & $W$ & $W_5$&  $m$	& \textbf{$\alpha_1$} & $h^c$ & $h^t$ & $A$ & $\mu$ \\ \midrule
\textbf{Computability}   & \cmark &  \cmark &\cmark &  \cmark & \cmark & \xmark & \xmark & \cmark & \cmark\\
\textbf{Units}   & \cmark &  \cmark &\cmark & \xmark & \xmark & \xmark & \xmark & \xmark & \xmark  \\
\textbf{Locality}   & \xmark &  \cmark & \cmark &\xmark & \xmark & \xmark & \xmark & \xmark & \xmark \\
\textbf{Constancy}   & \cmark &  \cmark &\cmark &\cmark & \cmark  & \xmark & \xmark & \xmark & \xmark\\
\textbf{End of career}   & \xmark &  \cmark &\cmark &\xmark & \xmark & \xmark & \xmark & \xmark & \xmark\\ 
\textbf{Packaging-independence}   & \cmark &  \cmark & \cmark &\xmark & \xmark & \xmark & \xmark & \xmark & \cmark\\

\hline
\end{tabular}
\end{table}

Based on the discussion above, we move on to consider only $W(t)$ and $W_5(t)$, which seem the most promising measures. 

\section{Case study of mathematical researchers} 
\label{ss:case}

We now discuss the performance of the measures on real data. 

\subsection*{Methods}
The experiments described below were carried out in February 2020. All raw data and computed data are available at ***insert link following publisher rules ***.

We concentrated for this article on the research area most familiar to us, namely mathematics. We compiled several datasets based on specific sets of researchers. The lack of availability of this data in an open format, or even in a proprietary one that allows for comprehensive analysis, is a major factor in hampering studies of this type. We resorted to making many time-intensive manual web-based queries. In order to test our hypotheses, for each researcher we require the total number of citations to her works in each calendar year. We used Web of Science Core Collection, which we chose because of its availability and reasonably wide coverage.

We extracted four sets of mathematicians and categorized them. The first set used consisted of all Abel Prize winners (with the exception of two for whom author name disambiguation was too difficult) and included 18 authors. The second consisted of 10 mathematicians from a single department (University of Massachusetts Amherst, Mathematics \& Statistics) with interests in a common subfield (algebraic geometry). The third dataset consisted of 10 authors generated ``randomly" from MathSciNet  (we chose authors of the most recent papers in algebraic geometry according to MathSciNet). The fourth consisted of all living winners of the Fields' Medal from 2006 to 2018 inclusive, and consisted of 13 authors. This gave a total of 51 mathematical researchers. For each researcher we took year 1 to be the first year $t$ for which $N(t) \geq 10$. 

Larger datasets would give more confidence in the results below, but they are clear enough to show that the axiomatically well founded measures $W$ and $W_5$ measure something about a researcher that enables us to distinguish between randomly chosen, successful and outstanding researchers.

\subsection*{Results} \label{ss:results}

\subsection{Variation of measures} 
\label{ss:variation}

\begin{figure}[htbp]
\includegraphics[width=15cm,height=11cm]{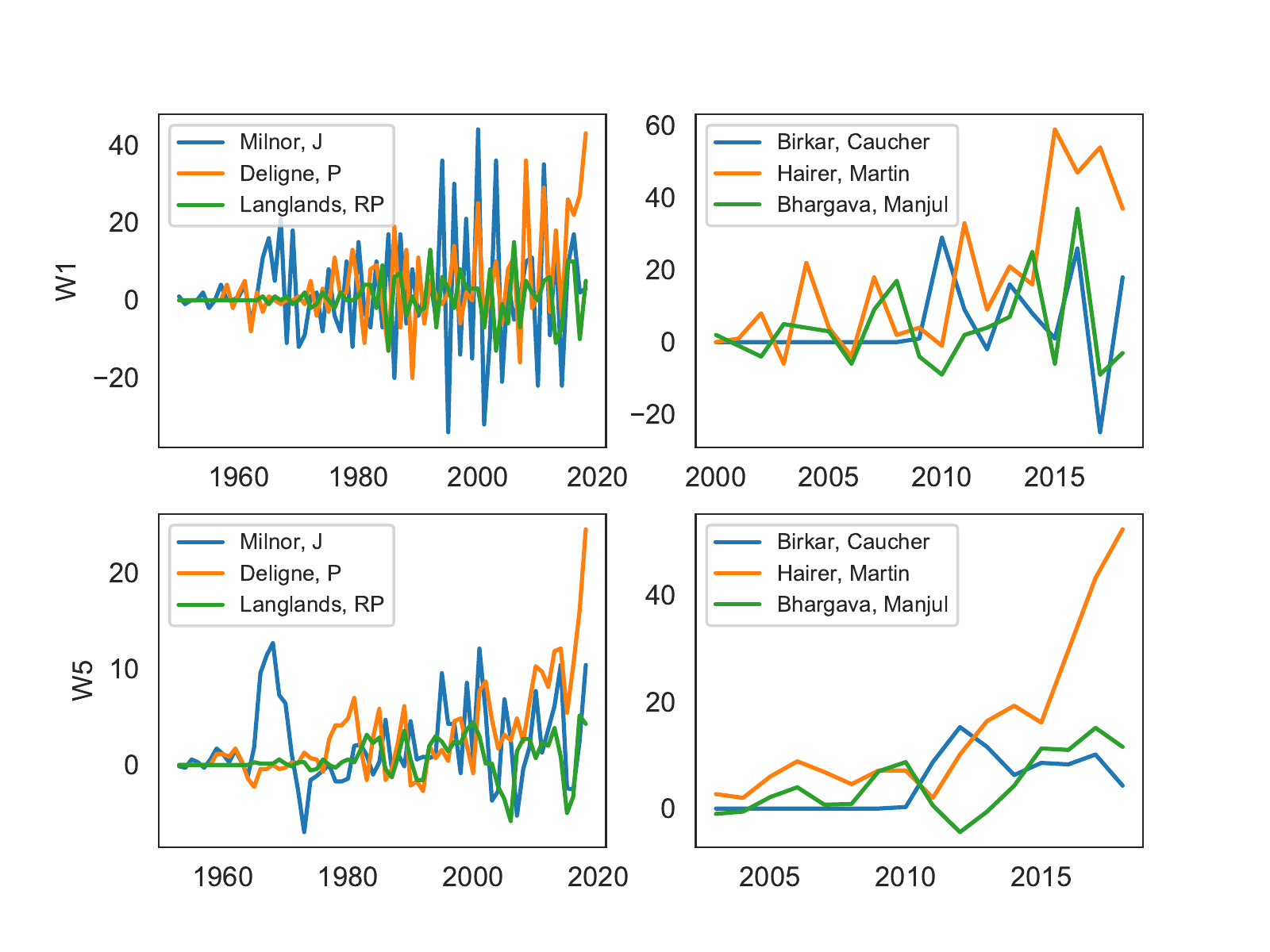}
\caption{Variation over time of values of $W$ and $W5$ for three selected Abel and Fields prizewinners}
\label{fig:variation-W-W5}
\end{figure}

In Figures~\ref{fig:variation-W-W5} we graph, for 3 Abel and 3 Fields prizewinners, $W(t)$ and $W_5(t)$ for $t$ from 5 years after career start (defined as the first year for which $N$ exceeds 10). 
As can be seen, there is substantial variation over time for each author. In Table~\ref{t:variation-abel+fields-W+W5} we give mean and standard variation of the values of the measures over the same time period.

\begin{table}[htbp]
    \begin{tabular}{lcccc}
    \toprule
    {} &  Mean(W) &    SD(W) &  Mean(W5) &   SD(W5) \\
    \midrule
    Serre, JP     &   1.0000 &   4.0452 &   -0.3636 &   2.0444 \\
    Atiyah, MF    &   9.2727 &  34.8792 &   10.7273 &  14.9522 \\
    Singer, IM    &   1.9091 &   4.2094 &    0.5584 &   1.1959 \\
    Lax, PD       &   5.9091 &  19.1095 &    5.1429 &   2.9410 \\
    Carleson, L.  &   1.4545 &   8.4892 &    1.1429 &   2.6719 \\
    Varadhan, SRS &   2.4545 &  27.8873 &    6.1039 &   5.8494 \\
    Tits, Jozef   &   1.0909 &   7.8097 &    0.8961 &   2.0066 \\
    Tate, JT      &   1.4545 &   6.3584 &   -0.4286 &   2.2686 \\
    Milnor, J     &   3.9091 &  11.0820 &    3.6364 &   5.9893 \\
    Szemeredi, E  &   2.4545 &   7.9357 &    2.1948 &   1.2383 \\
    Deligne, P    &   0.0909 &   2.3914 &   -0.0779 &   0.8729 \\
    Sinai, YG     &   0.5455 &   2.4629 &    0.1688 &   0.5963 \\
    Nash, JF      &   0.6364 &   4.7725 &    0.5195 &   1.6551 \\
    Nirenberg, L  &   3.3636 &   8.7102 &    2.9481 &   1.8579 \\
    Wiles, A.     &   2.7273 &   8.4325 &    1.0909 &   2.2317 \\
    Meyer, Yves   &   9.6364 &  12.1676 &    5.9351 &   5.6300 \\
    Langlands, RP &   0.8182 &   5.4242 &    0.9870 &   1.2098 \\
    Uhlenbeck, K  &   2.8182 &  18.6977 &    5.7792 &   5.4519 \\
    \bottomrule
    \end{tabular}
    \caption{Mean and standard deviation for $W$ and $W_5$ for Abel prizewinners}
    \label{t:variation-abel+fields-W+W5}
\end{table}

\subsection{Predictive value of measures} 
\label{ss:predict}
As explained by Penner et al. \cite{penner2013predictability}, cumulative increasing measures such as the $h$-index contain intrinsic autocorrelation, which vastly overstates their predictive power. They find that the actual ability of the $h$-index to predict future citations from future publications is rather low. In our case, we are dealing with noncumulative measures, whose predictive power is not so clear. The results of Section~\ref{ss:variation} show considerable variation in the year-to-year values of $W(t)$ and $W_5(t)$, so the idea of the simple model, that these measures are constant and hence precisely determine something intrinsic to the researcher, is not plausible. We do not expect to be able to predict the value of $W(t+5)$, for example, from $W(t)$ only. However, the results in Section~\ref{ss:expert} show that gross distinctions between researchers at different levels of impact can be made (when dealing with researchers in the same fairly narrowly defined field), and these seem to mean something.

\begin{figure}[htbp]
\includegraphics[width=8cm,height=6cm]{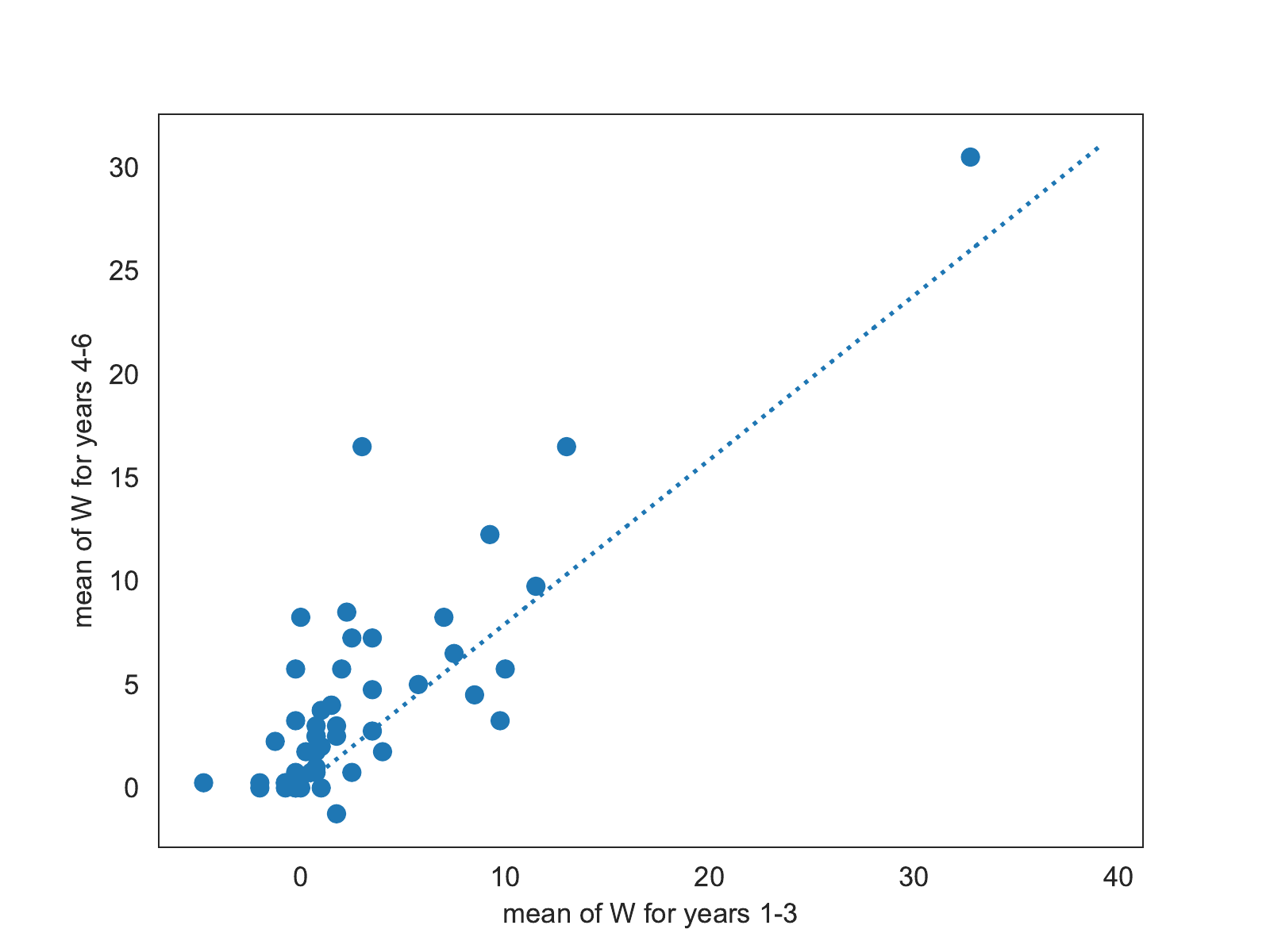}
\caption{Regression of mean of $W$ in years 4-6 against mean in years 1-3}
\label{fig:regress}
\end{figure}

In order to obtain a better idea of predictive power, for the union of our datasets we computed the mean in years 3--5 of career of $W$, and used this to attempt to predict the mean in years 6--8 of the same measure. The ordinary least squares linear regression results are displayed in Figure~\ref{fig:regress} (note that the extreme outlier Terence Tao was removed from the dataset, as was the very young Peter Scholze leaving 49 researchers). The value $R^2 = 0.744$ shows a high level of predictive power. Note that the definition of $W$ means that we are trying to predict
a linear combination of $N(8), N(7), N(5)$ and $N(4)$  from a linear combination (with the same coefficients) of $N(5), N(4), N(2), N(1)$, and there is no a priori reason to expect this to have such a high coefficient of determination.

\subsection{Relation to expert judgment}
\label{ss:expert}

\begin{figure}[htbp]
\includegraphics[width=12cm,height=9cm]{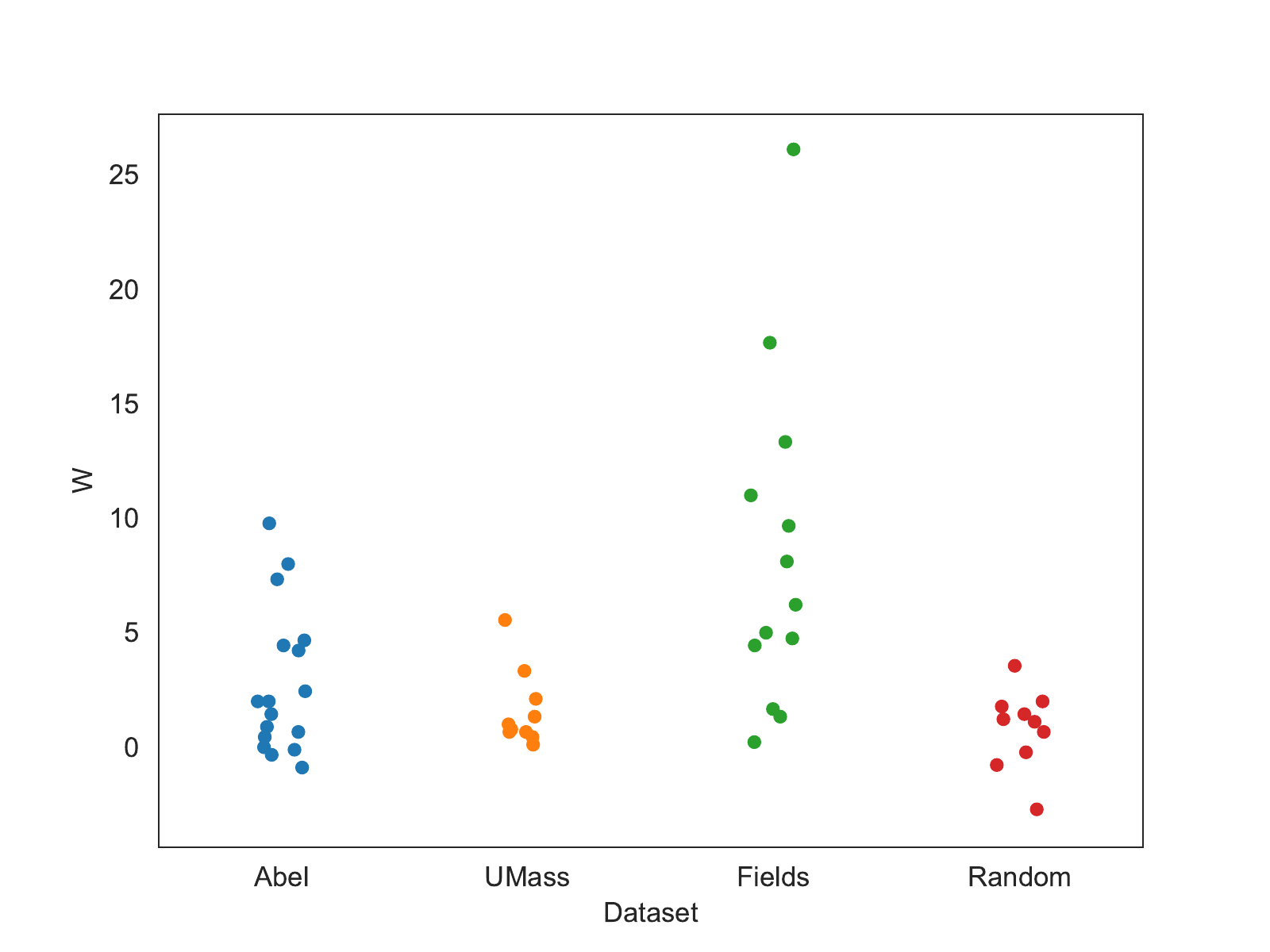}
\caption{Scatterplot of values of $W$ for four different datasets}
\label{fig:scatterplot}
\end{figure}

A priori, we expect that the citation acceleration for randomly chosen authors should be lower overall than that of the UMass researchers. Also, we expect the Fields dataset and Abel dataset to have higher values of the measures than the UMass dataset. Given the age of the members of the Abel dataset, we expect a small advantage to the Fields dataset. 
All these are borne out by the results shown in Figure~\ref{fig:scatterplot}.

\section{Discussion} 
\label{s:con}

\subsection*{Limitations}

Before summarizing our main points, we acknowledge that the entire subject of bibliometrics is suspect in the eyes of many researchers, because it has been excessive relied on by unimaginative bureaucrats having influence over researcher careers. We share these concerns, and the present article was motivated by the idea that if we researchers are forced to be evaluated by simple metrics, we can at least have some agency in their design. The popularity of the $h$-index, for example, is very mysterious to us, given its weak theoretical foundations \cite{waltman2012inconsistency}. Any one measure can be strategically manipulated, but we hope that by use of sufficiently many metrics with low theoretical correlation, incentives for researchers to act in ways that are not helpful to science overall will be reduced.

All citation measures in the literature are susceptible to many problems (including missing data, author name disambiguation, negative citations, contributions of multiple authors, citation inflation owing to growth in number of researchers). Also, there is the problem of  normalization across different fields (for example, one citation in mathematics corresponds roughly to 19 in physics and 78 in biomedicine \cite{podlubny2005comparison}). This question of ``field" is a difficult one - it is obvious that certain areas of mathematics have communities of different sizes, leading to substantial variation in the number of citations across areas. Thus, as usual, there is no completely automated substitute for human judgment. 

\subsection*{Positive outcomes}

The index $W(t)$ introduced in this article seems to measure something specific to a researcher that is related to their recent productivity and impact, and seems promising as a way to make coarse distinctions between researchers in the same field who may be at different career stages. It behaves well with respect to natural axioms. It seems fairly well correlated with subjective measures of research impact or quality. It is less sensitive to the way in which ideas are packaged into individual publications, and considerably easier to compute, than the $m$-index (under the assumption that splitting a paper splits the citations in the obvious way, the $m$-index discourages extreme ``salami-slicing", whereas $W$ is indifferent to it and $P$ encourages it).

Insofar as citation metrics are to be increasingly used for evaluation of researchers and especially for allocation of resources to them, the $W$-index provides another useful (perhaps the single most useful found so far) measure of recent publication activity leading to citation impact, and one that has decent predictive value. 

\subsection*{Future work}
\label{ss:future}
Bouyssou and Marchant \cite{bouyssou2014axiomatic} state that their paper explicitly does not deal with any indicators intended to adjust for career age, and the last part of the paper suggests further work in such a direction. We offer the present work as an initial contribution, and intend to follow up. A stream of research initiated by Woeginger \cite{woeginger2008axiomatic} deals with axiomatic \emph{characterizations} of the $h$-index --- that is, a set of axioms which taken together uniquely determine the $h$-index. Our experience with characterization theorems (and also impossibility theorems where ``too many" axioms are chosen, a prominent approach in social choice theory, for example)  is that very often the axiom systems consist of a few innocuous assumptions and one that is much less intuitive and essentially encodes the desired result. Nevertheless, it would be interesting to obtain an axiomatic characterization of our measure $W$, for example. 

The relation \eqref{eq:3} derived from the simple model has wider validity than might be expected at first sight.
Mannella \& Rossi \cite {mannella2013time} find via a study of 1400 Italian physicists that this quadratic relationship holds well on real data, and empirically find the best fit value of $\beta= 0.53$  in \eqref{eq:3}, agreeing with the rough calculations of Hirsch based on a smaller dataset of physicists. Yong \cite{yong2014} showed analytically, based on theory of 
random partitions, that a very good estimate should be $\beta = \sqrt{6}\ln 2/\pi\approx 0.54$. He also demonstrated the accuracy of this approximation on a small dataset of prominent mathematicians.

However, Mannella \& Rossi also showed that the time scaled index 
\begin{equation}\label{eq:5}
\alpha_2(t)=\frac{h(t)}{\sqrt{t}}
\end{equation}
is approximately time independent on their dataset. This implies that the number of citations is better described as linear rather than quadratic, which is clearly inconsistent with the simple model. Attempts to estimate citation acceleration depend on the quadratic growth of citations with time, so a possible linear relationship will have an effect on the measures used in this paper: in that case, the acceleration would be identically zero. We feel that the growth of citations by an author with time should be studied seriously on much larger datasets than we have treated here.

In order to concentrate on the main concept of citation acceleration, we have omitted more subtle issues such as  weights for coauthored papers and normalization by the size of the research field. These of course could be explored.


\bibliographystyle{plain}

\bibliography{refs}

\end{document}